\documentclass[journal]{IEEEtran}
\usepackage{CJK}
\usepackage{amssymb}
\usepackage{amsmath}
\usepackage{graphicx, subfigure}
\usepackage{url, cite}
\usepackage{verbatim}
\usepackage{booktabs}
\usepackage{multirow}
\usepackage{bigstrut}
\usepackage{booktabs}
\usepackage{amsmath,cite,graphicx,times,subfigure,url,verbatim}
\usepackage{textcomp}
\usepackage{setspace}
\usepackage{epstopdf}

\newtheorem{subsec:coding}{subsec:coding}

\usepackage{color}

\begin{document}


\title{Urban Healthcare Big Data System based on Crowdsourced and Cloud-based Air Quality Indicators}

\author{
Min~Chen,~Jun~Yang,~Long~Hu,~M.~Shamim~Hossain, Ghulam~Muhammad
\thanks{M. Chen, J. Yang and L. Hu are with School of Computer Science and Technology, Huazhong University of Science and Technology, Wuhan 430074, China.}
\thanks{M. Chen, J. Yang and L. Hu are also with Embedded and Pervasive Computing (EPIC) Lab, Huazhong University of Science and Technology, Wuhan 430074, China.}
\thanks{M. Shamim Hossain is with the Department of Software Engineering, College of Computer and Information Sciences, King Saud University, Riyadh 11543, Saudi Arabia.}
\thanks{G. Muhammad is with the Department of Computer Engineering, College of Computer and Information Sciences, King Saud University, Riyadh 11543, Saudi Arabia.}
}

\markboth{Accepted: IEEE COMMUNICATIONS MAGAZINE, VOL. XX, NO. YY, MONTH 2018}{}

\maketitle


\begin{abstract}
The ever-accelerated process of globalization enables more than half the population to live in cities, thus the air quality in cities exerts critical influence on the health status of more and more urban residents. In this article, based on the urban air quality data collected through meteorological sites, mobile crowdsourcing and IoT sensing, along with user's body signals, we propose an urban healthcare big data system, named UH-BigDataSys. In this article, we first introduce a method of integrating multi-source air quality data, for the data preparation of the artificial intelligence based smart urban services. Then, a testbed of UH-BigDataSys is set up with the deployment of air quality-aware healthcare applications. Finally, we provide health guidance for urban residents in aspects of respiratory diseases, outdoor travel, and sleep quality, etc. The ultimate goal of UH-BigDataSys is to lead urban resident a more healthy life.
\end{abstract}


\section{Introduction}
\begin{table*}
  \renewcommand{\arraystretch}{1.3}
  \newcommand{\tabincell}[2]{\begin{tabular}{@{}#1@{}}#2\end{tabular}}
  \caption{Profile of Sensing Data}
  \begin {center}
  \begin {tabular}{|l|l|l|l|l|l|l|l|l|l|l|}
  \hline
\textbf{Data Category}       & \multicolumn{3}{l|}{\textbf{Item}}       & \multicolumn{5}{l|}{\textbf{Typical Value  }}  & \multicolumn{2}{l|}{\textbf{Nullable  }}\\
\hline
\multirow{9}*{Air Quality Data} & \multicolumn{3}{l|}{Location(Longitude)}  &  \multicolumn{5}{l|}{114.3672}                &  \multicolumn{2}{l|}{No} \\
                          & \multicolumn{3}{l|}{Location(Latitude)}         &  \multicolumn{5}{l|}{30.5719 }                &  \multicolumn{2}{l|}{No} \\
                          & \multicolumn{3}{l|}{Time}                       &  \multicolumn{5}{l|}{2017-05-15 12:30:01}     &  \multicolumn{2}{l|}{No} \\
                          & \multicolumn{3}{l|}{PM2.5}                      &  \multicolumn{5}{l|}{29 μg/m3 }               &  \multicolumn{2}{l|}{Yes} \\
                          & \multicolumn{3}{l|}{PM10 }                      &  \multicolumn{5}{l|}{- μg/m3  }               &  \multicolumn{2}{l|}{Yes} \\
                          & \multicolumn{3}{l|}{CO}                         &  \multicolumn{5}{l|}{1.3 mg/m3}               &  \multicolumn{2}{l|}{Yes} \\
                          & \multicolumn{3}{l|}{NO2}                        &  \multicolumn{5}{l|}{25 μg/m3 }               &  \multicolumn{2}{l|}{Yes} \\
                          & \multicolumn{3}{l|}{O3 }                        &  \multicolumn{5}{l|}{180 μg/m3 }              &  \multicolumn{2}{l|}{Yes} \\
                          & \multicolumn{3}{l|}{SO2 }                       &  \multicolumn{5}{l|}{14 μg/m3  }              &  \multicolumn{2}{l|}{Yes} \\
\hline
\multirow{8}*{Physiological Data}        & \multicolumn{3}{l|}{Location(Longitude)}        &  \multicolumn{5}{l|}{114.3894}                &  \multicolumn{2}{l|}{No} \\
                          & \multicolumn{3}{l|}{Location(Latitude)}         &  \multicolumn{5}{l|}{30.4822 }                &  \multicolumn{2}{l|}{No} \\
                          & \multicolumn{3}{l|}{Time               }        &  \multicolumn{5}{l|}{2017-05-15 09:21:45}     &  \multicolumn{2}{l|}{No} \\
                          & \multicolumn{3}{l|}{ECG                 }       &  \multicolumn{5}{l|}{1221(ADC Sampling)}      &  \multicolumn{2}{l|}{Yes} \\
                          & \multicolumn{3}{l|}{EMG                  }      &  \multicolumn{5}{l|}{2542(ADC Sampling)}      &  \multicolumn{2}{l|}{Yes} \\
                          & \multicolumn{3}{l|}{Heart Rate            }     &  \multicolumn{5}{l|}{74 (beats per minute) }  &  \multicolumn{2}{l|}{Yes} \\
                          & \multicolumn{3}{l|}{Body Temperature       }    &  \multicolumn{5}{l|}{29 \textcelsius{} }      &  \multicolumn{2}{l|}{Yes} \\
                          & \multicolumn{3}{l|}{Blood oxygen            }   &  \multicolumn{5}{l|}{98 \%   }                &  \multicolumn{2}{l|}{Yes} \\
\hline
\multirow{5}*{M-AQI} & \textbf{(Longitude} & \textbf{Latitude)} & \textbf{Time} & \textbf{AQI} & \textbf{PM2.5} & \textbf{PM10}  & \textbf{CO} & \textbf{NO2}  & \textbf{O3 }  & \textbf{SO2}\\ \cline{2-11}
                        &114.3672 &  30.5719    & 2017-05-15(12:00-13:00)      & 75      & 29       & 60    & 1.3      & 25      & 180       & 14 \\  \cline{2-11}
                        &114.2511 & 30.5514     & 2017-05-15(12:00-13:00)      & 70      & 9        & 90    & 0.5      & 30      & 144       & 7 \\   \cline{2-11}
                        &114.2836 & 30.6197     & 2017-05-15(12:00-13:00)      & 52      & 9        & 53    & 0.5      & 21      & 141       & 8 \\   \cline{2-11}
                        &114.3006 & 30.5494     & 2017-05-15(12:00-13:00)      & 35      & 10       & 22    & 0.6      & 24      & 112       & 10 \\  \cline{2-11}
\hline
    \end {tabular}
 \label{tab:SensingData}
 \end {center}
\end {table*}

In the 2016 edition of World Health Statistics, the World Health Organization (WHO) reported that people died of diseases related to air pollution reached 6,500,000 in 2012. And this number exceeds 11\% of death toll in 2012. Meantime, it is explicitly pointed out by WHO in report ``Ambient (outdoor) air quality and health'' that air pollution is a main environmental risk that exerts influence on health. The diseases such as stroke, heart disease, lung cancer, chronic and acute respiratory illnesses~\cite{disease} may be prevented by decreasing air pollution level. The lower the air pollution level is, the healthier the cardiovascular and respiratory systems of people are, no matter whether it is in long term or in short term. The impact of air pollution on health of human in urban environment is presented in~\cite{1,2}. Thus, air quality evaluation can bring significant impact on health status of the enormous urban residents.

Air Quality Indicators (AQIs) include an internationally-used parameter set to evaluate air quality, and the statistics reflects five pollution standards, including ground-level ozone, particulate matter, carbonic oxide, sulfur dioxide and nitrogen dioxide. At present, AQIs are usually measured by weather monitoring sites. Due to their high construction cost, limited coverage and insufficient quantity, the air quality data collected by traditional meteorological sites are not enough to portrait the real situations. Fortunately, with the evergrowing number of smart devices and mobile terminals, urban residents with portable mobile devices can contribute to sense ambient air quality in real time~\cite{7,8}. Then, the data are stored and shared in clouds. Thus, the public can participate in the collection of urban air quality data through mobile devices and the perception of urban air quality may be established over large scale networks~\cite{4}. This crowdsourcing methods for urban environment sensing are discussed in~\cite{5,6}.
Furthermore, with the technology advances regarding Internet of Things (IoT) and vehicle networking~\cite{vehicle}, smart building and vehicles are also equipped with monitoring facilities. The related design of hardware, software as well as architecture for urban environment sensing are presented~\cite{11,13}. Zheng et. al~\cite{14} propose to predict air quality by analyzing the correlation among air quality data collected by weather monitoring and vehicles.
Through the above methods, more comprehensive analytics is applied with and more abundant urban air quality data, which are acquired with lower cost. The seamless integration between public daily life and the perception of air quality can be realized based on the sensing assisted by mobile users. The purpose is not only to improve the coverage area and the efficiency of air quality data collections, but also enable the efficient processing and sharing of real-time data flows in combination with the historical information.

However, to the best of our knowledge, there is no work considering the integration of meteorological site data, mobile device crowdsourced data and IoT sensing data to improve the accuracy of AQI analysis. Furthermore, user's physiological indicators are not considered in the solution.
In the face of abominable urban air quality conditions in lots of developing countries, how to conduct joint urban air quality sensing and health status analytics is critical to provide personalized health monitoring solution. Although challenging, it brings extensive social value to diagnose individual physiological state based on urban air quality data for improving the quality of people's life. Thus, an innovative healthcare solution should be discussed based on user-oriented physiological data and urban AQI.

In this article, we propose urban healthcare big data system (i.e., UH-BigDataSys), where data integration and physiological indicator are considered. Meteorological sites, mobile crowdsourcing and IoT sensing are adopted for air quality data collection, in order to provide urban residents with more comprehensive and more accurate air quality monitoring services. Besides air quality monitoring, realtime physiological index monitoring for user is also realized by the use of wearable devices. Corresponding guidance for health and daily activities is provided to urban residents dynamically as air quality changes. As for a user who participates outdoor activities, personalized and healthy activity plan is recommended by UH-BigDataSys, with the analytics of his or her current physiological status and air quality in his or her surrounding environment. The introduction of air quality index provides health analysis for the user with more abundant dimensionalities of information and improves efficiency in health guidance. The main contributions of this article are as follows:
\begin{itemize}
  \item M-AQI (Multidimensional Air Quality Indicators) big data integration method is proposed based on AQI sensing in three networking levels. First, crowdsourced AQI data is collected as the first level. Then, data fusion of AQI data is considered in edge cloud-based level. Finally, AQI data are integrated on remote cloud or meteorological supercomputing platform.
  \item An innovative healthcare monitoring system via urban big data (i.e., UH-BigDataSys) is proposed based on M-AQI big data and physiological data of user. Compared with traditional health monitoring system on basis of physiological indices, the range of applications for UH-BigDataSys is broader. UH-BigDataSys is not only limited to traditional health monitoring, it can also give suggestions in a higher level in combination with current health status of user and their surrounding environment, for example, the travel guidance to patients with respiratory diseases may be personalized by this system.
  \item The demonstration application platform for UH-BigDataSys is established. Typical applications include early forcasting for diseases related with urban air quality, travel guidance to patients with respiratory diseases, and monitoring for user emotion and sleep quality related to indoor air quality.
\end{itemize}

The remain contents are arranged as follows in this article. Section II introduces the method for the integration of M-AQI big data, Section III introduces the design of UH-BigDataSys, Section IV discusses the testbed of UH-BigDataSys, finally Section V gives conclusion.

\begin{figure}
\centering
\includegraphics[width=3.5in]{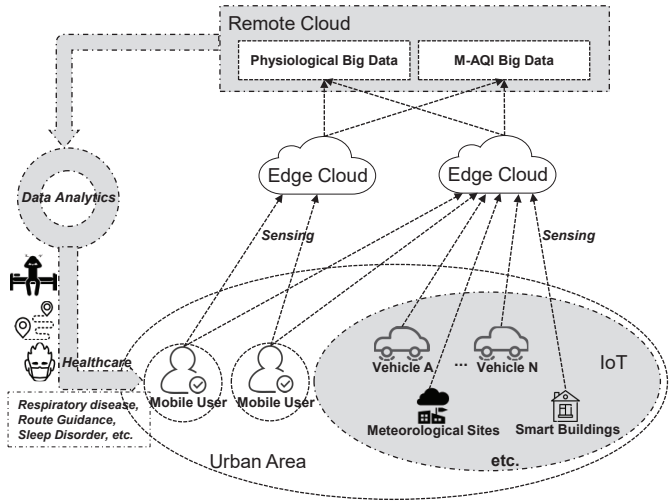}
\caption{Flowchart of M-AQI Big Data Integration}
\label{fig001}
\end{figure}

\section{M-AQI Big Data Integration based on Crowdsourcing and Edge Clouds}

In this section, we mainly consider two categories of smart city data related with the health status of urban residents, i.e., AQI data and user's body signals. To collect AQI data, we use various AQI sensing devices which are widely spread on urban environments~\cite{15}. For the collection of body signals, wearable 2.0 device (e.g., smart clothing) can be utilized. Fig.~\ref{fig001} shows the procedure of the integration of M-AQI big data. The heterogeneous AQI data are integrated for data fusion and sharing in edge clouds and data center cloud. With the accumulation of M-AQI data, the related big data analytics is conducted in the clouds. This section will address issues on sensing and fusion process of AQI data.

\subsection{AQI Data Sensing through Crowdsourcing}

In Table~\ref{tab:SensingData}, three categories of data are classified. There are 6 fundamental indicators in the category of ``Air Quality Data'', i.e., PM2.5, PM10, CO, NO2, O3 and SO2. By the use of mobile devices carried by urban residents, the technique of crowdsourcing is useful for sensing AQI data. With various advantages such as low cost, large sensing coverage, location-aware and personalized data collection, it can be widely used in various aspects of daily life, such as traffic, environment monitoring and healthcare. However, this method also exhibits some disadvantages such as poor data quality and intermittent data provisioning.

AQI sensing via crowdsourcing requires to distribute numerous data sensing tasks to available caching and computing resources of mobile devices carried by urban residents. The specific steps of crowdsourcing are list as follows. Firstly, UH-BigDataSys should establish an association matrix using mobility data of urban residents such as GPS, acceleration and moving speed, etc. Based on the association matrix, mobility pattern and movement features of the residents are extracted. Then, we can establish the residents' behavioral model. Finally, optimal selection can be achieved regarding those urban residents whose routes of mobility match with the sensing points of AQI. Furthermore, UH-BigDataSys should choose those urban residents with higher credits (i.e., users who contribute high-quality sensing data, as data quality evaluated by accuracy, timeliness, correlativity and integrity, etc.) and assigns data sensing task to them.

\subsection{Edge Cloud-based AQI Integration}

\begin{figure}
\centering
\includegraphics[width=3in]{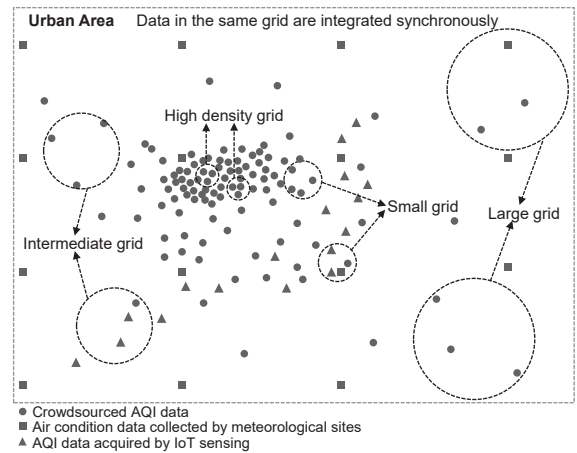}
\caption{Schematic diagram for dynamic data integration based on time-space characteristic and data density distribution}
\label{fig002}
\end{figure}

As shown in Fig.~\ref{fig002}, there are three categories of AQI data collected: (i) AQI data sensed by urban residents via crowdsourcing; (ii) AQI monitored by IoT sensing with monitoring facilities (such as vehicles and intelligent buildings, etc.); (iii)AQI data acquired in meteorological sites deployed by meteorological agency. These data are synchronized to edge clouds through various communication modes. However, the diversity of data source causes heterogeneity of AQI data in time-space characteristic, data density distribution and data accuracy. Heterogeneity of data density refers to the different densities of AQI data in different time slots or different regions. The heterogeneity of data accuracy refers to the accuracies for data from different sources are different. Intuitively, the accuracy of AQI data from meteorological monitoring sites is highest. In comparison, AQI data acquired from mobile devices are low, especially under their high-speed movements. Thus, how to integrate multi-source and multi-quality AQI data is a challenging problem.

In this section, we propose an edge cloud-based AQI integration method and it can be conducted on edge cloud and remote cloud synchronously. First, we extracte location (i.e., longitude and latitude) feature and time feature of AQI data, as shown in Table~\ref{tab:SensingData}. Then, we partition those data based on time slot feature and determine the size of grid on basis of data density in a single time slot. The grid size is set to be smaller if the data density in that region is higher, thus to guarantee the grid with finer granularity in those regions with high data density. Typically, those regions correspond to residential areas with high dense population. Thus, it needs to provide precise data with finer granularity. Finally, for the heterogeneity of data accuracy, we use weighted average for the sake of simplicity,
\begin{equation}
\begin{aligned}
v&=\frac{w_1}{m}\sum_{i=1}^{m}v_i+\frac{w_2}{n}\sum_{i=1}^{n}v_i+\frac{w_3}{l}\sum_{i=1}^{l}v_i
\end{aligned}
\end{equation}
where $m$ represents the number of data segments in a certain time slot through crowdsourcing, $n$ stands for the number of data segments collected from meteorological sites, $l$ stands for the number of data segments via IoT sensing, $v_i$ stands for the sampling value of an attribute (e.g., PM2.5), where $w_1$, $w_2$ and $w_3$ denote the weights of various measurements, and $w_1+w_2+w_3=1$. After the calculation is completed, combination for each attribute value in the same category in a certain time slot is conducted to generate a new M-AQI record. And the profile of M-AQI data is shown in column ``M-AQI" in Table~\ref{tab:SensingData}, thereinto, the value of AQI will be obtained by converting the 6 air quality attributes.

\subsection{AQI Integration on Remote Cloud or Supercomputing Platform}

M-AQI data produced at edge clouds will be synchronized and shared to remote cloud. Thus data integration can also be conducted at remote cloud with same method as at edge clouds.
There are stringent requirements in aspect of computing and storage for M-AQI big data applications. With limited resources, edge clouds can not fulfill the requirements of authentic big data applications. With its particular performance in terms of convenience, scalability and on-demand services, remote cloud is able to provide ultimate guarantee for M-AQI big data analysis.

Furthermore, supercomputing platform has massive high-performance computing resource, which provides new possibility for expanding the capability of remote cloud.
Because of the deficiency of user-friendly interface, current supercomputing platform can not provide convenient and interactive services.
By the construction of virtual resource pool, uniform management of supercomputing resources may be formed, and on basis of which connectivity is convenient with affordable cost in terms of supercomputing platform or remote cloud.
Furthermore, uniform management of supercomputing resources could define virtualized interface, protocol and software module for supercomputing platform, to abstract resources of supercomputing platform into open services. Finally, the common user-oriented supercomputing resource services can be obtained, and on basis of which, M-AQI big data services may provide urban residents with more accurate and personalized services.

\section{Design of UH-BigDataSys}
\begin{figure*}
\centering
\includegraphics[width=6in]{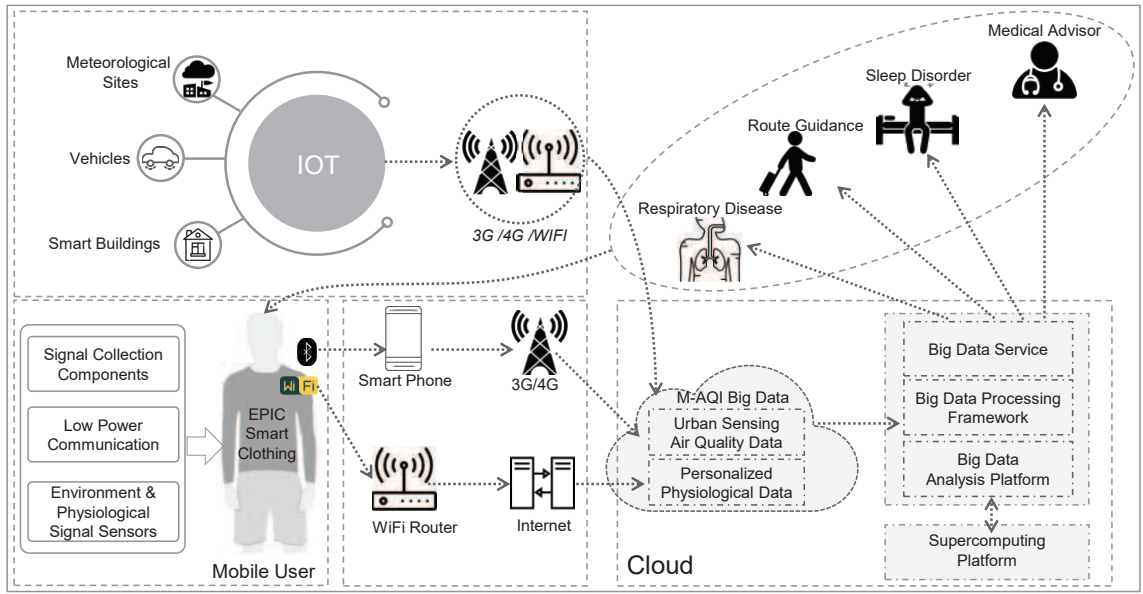}
\caption{System Architecture of UH-BigDataSys}
\label{fig003}
\end{figure*}

The architecture of UH-BigDataSys is shown in Fig.~\ref{fig003}, based on the acquired M-AQI big data, UH-BigDataSys also collects personalized physiological data of user via wearable 2.0 devices (e.g., smart clothing). Furthermore, the system jointly analyze physiological big data of the user and M-AQI big data based on user's location information. Finally, it provides the user with health advices for respiratory diseases, outdoor travel, and sleep quality control, etc.

For physiological big data collection, smart clothing is a great choice for urban residents under various scenes~\cite{wearable2}. It integrates textile clothing and body sensors, and exhibits good performance in term of sensor deployment, user's comfortableness, and low-power communications. The physiological data acquired by smart clothing are shown in Table~\ref{tab:SensingData} which include 5 fundamental indicators, i.e., electrocardiograph (ECG), electromyography (EMG), heart rate, saturation of blood oxygen and body temperature. The ECG and EMG are acquired via textile dry electrodes in smart clothing. Heart rate can be figured out from original ECG signals. As for the saturation of blood oxygen, optical sensor may be integrated into smart clothing to achieve non-invasive detection. Body temperature can be acquired by NTC thermistor sensor. Finally, data processing modules in smart clothing receive and process original signals generated by each sensor, then convert them into digital signals, and store them into its memory. Local processing unit determines whether to send physiological data to intelligent terminals or not according to status of communication module. Typically, intelligent terminal is a smart phone, for further transmitted them to cloud, thus to finally realize persistent storage for physiological data of the user in cloud.

\begin{figure*}
\centering
\subfigure[]{\label{fig04b} \includegraphics[width=1.3in]{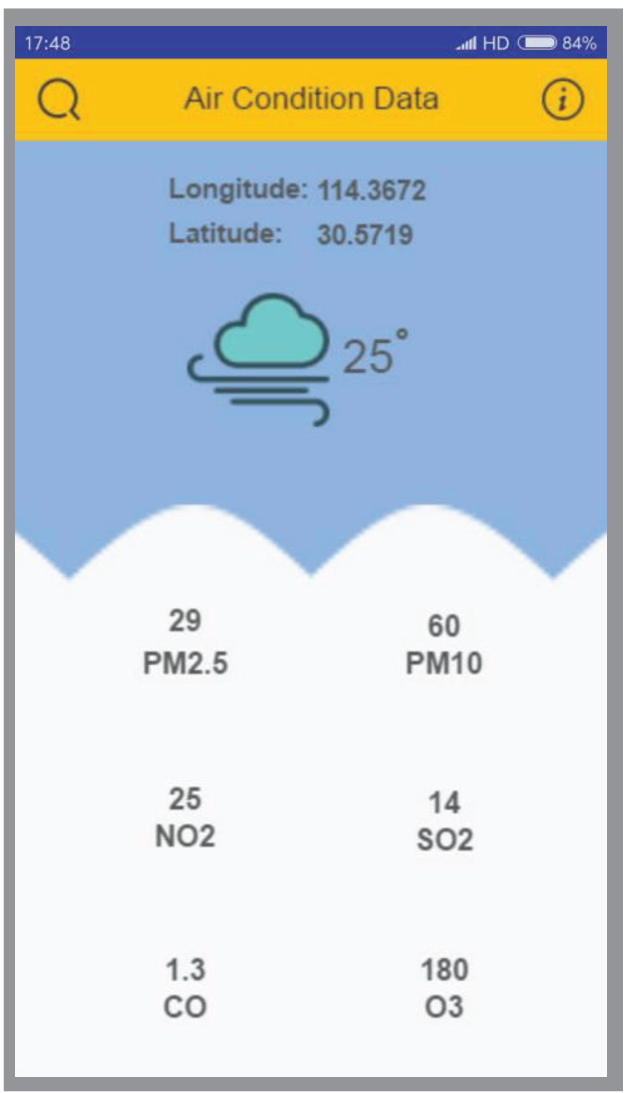}}
\subfigure[]{\label{fig04c} \includegraphics[width=1.3in]{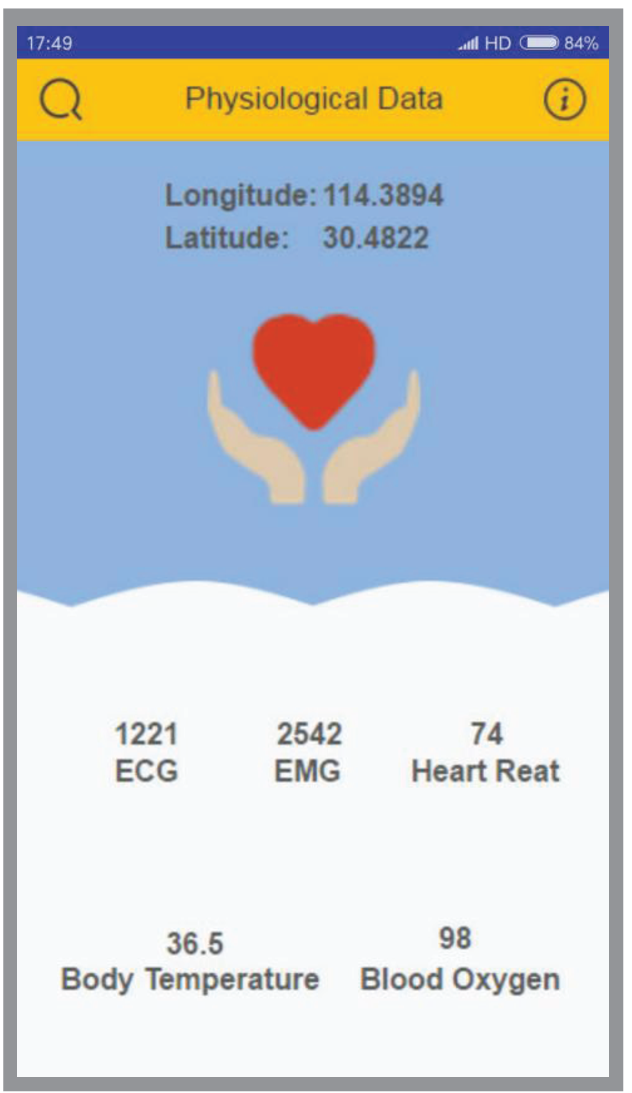}}
\subfigure[]{\label{fig04d} \includegraphics[width=1.3in]{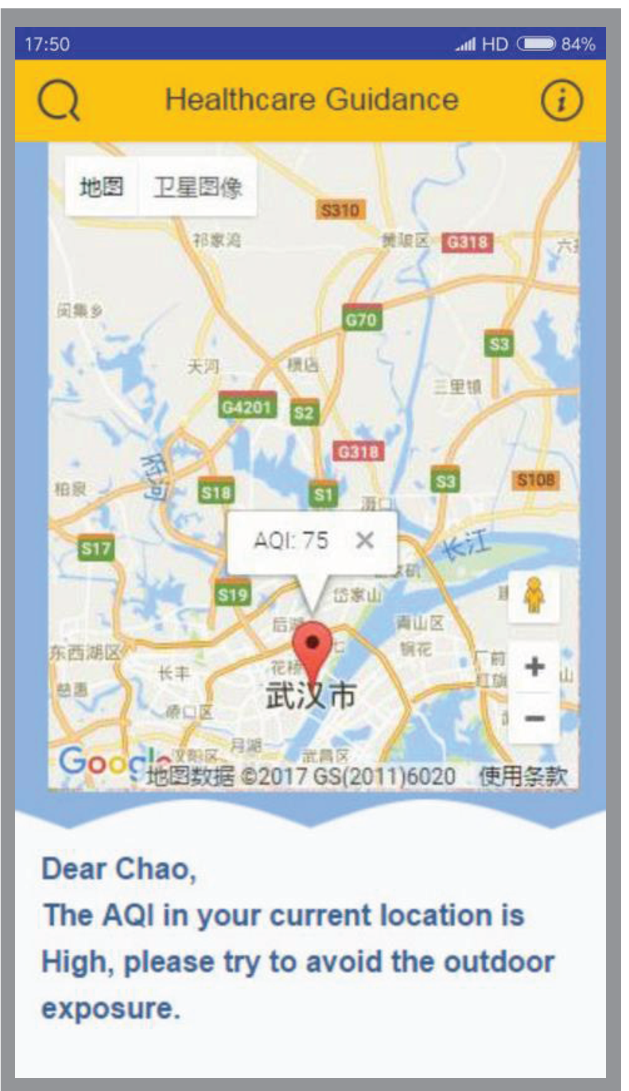}}
\subfigure[]{\label{fig04a} \includegraphics[width=2.7in]{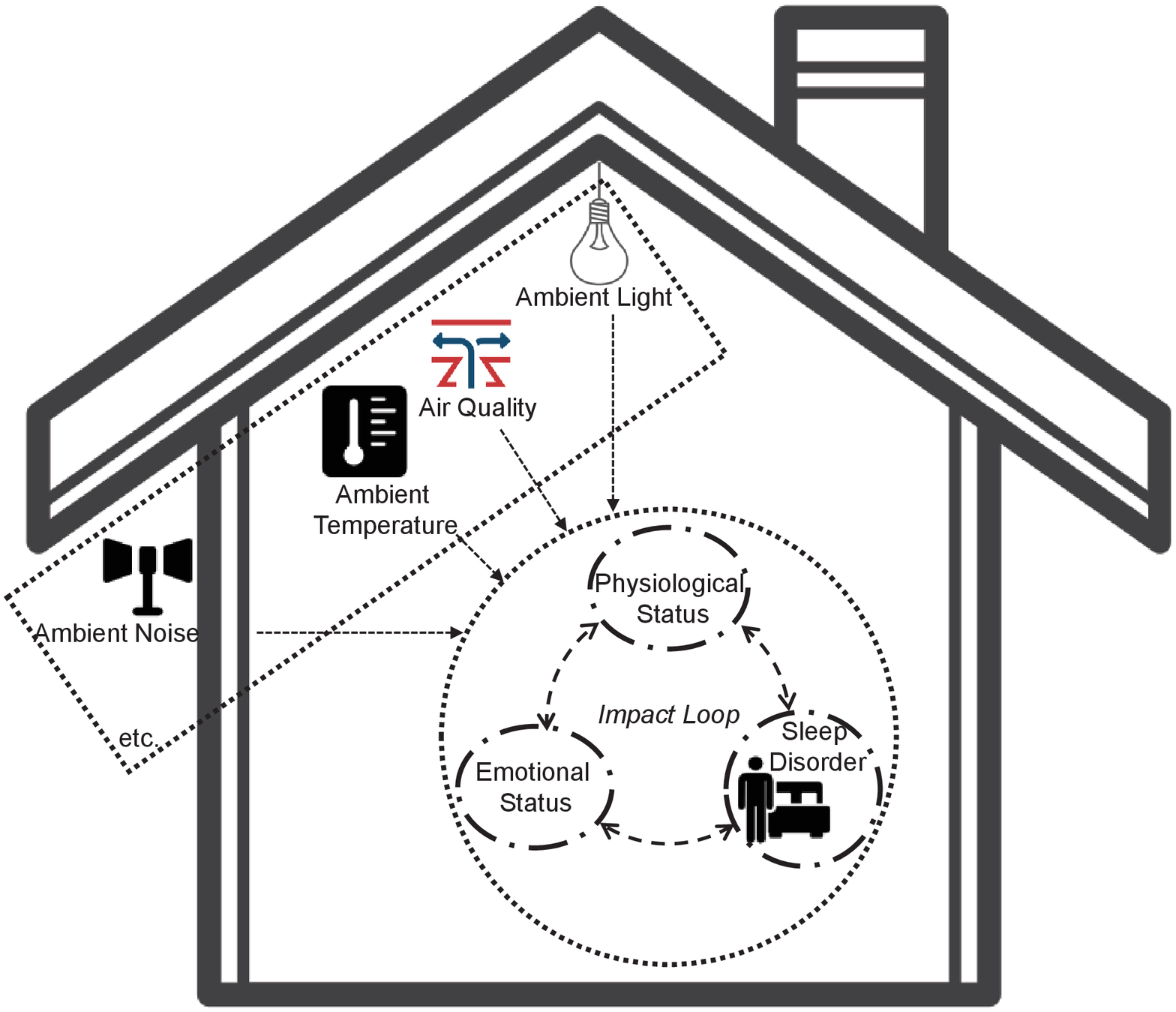}}
\caption{ (\textbf{a}) Air Condition Data, (\textbf{b}) Physiological Data, (\textbf{c}) Healthcare Guidance, \textbf{d}) Impact of Sleep Disorder.}
\label{fig004}
\end{figure*}

With the sustainable sensing and analysis on physiological data of urban residents, UH-BigDataSys can provide guidance based on physiological and mental status of the user. Furthermore, through the detection on air quality, UH-BigDataSys enriches the perception of surrounding environment of urban residents. Finally, UH-BigDataSys can provide health guidance to urban residents, with the combination of physiological data and air quality data around them. For example, in outdoor environment, a resident with physiological diseases, especially respiratory disease, will get timely warning from UH-BigDataSys about air quality around him or her, and remind him or her to pay attention to air quality conditions, thus to avoid deterioration of his or her respiratory disease. Meantime, through sensing on air quality data widely and dynamically in the city, UH-BigDataSys will also advise user for outdoor activities, plan the outdoor route for user when necessary, and guide user to avoid epidemic area or area with high air pollution. As for the indoor environments, air quality would also exert influence on disease and emotion of humans\~cite{emotion}. For example, the dreary air would bring discomfort to human body and exert influence on sleep of residents at night. Based on physiological data of user and indoor air quality data, UH-BigDataSys extract the correlations among sleep, daily activities, air quality and health of user.
Furthermore, on the basis of analyzing the correlations among all factors, UH-BigDataSys constructs measure model to figure out the influence of air quality on sleep status of user.

\section{A Testbed for UH-BigDataSys}
We have deployed a testbed to evaluate the performance of UH-BigDataSys system, the infrastructure we design for our testbed is a minimal cloud platform, which consists of 4 different servers, i.e., one controller, one networker as well as two computing nodes. And the executive environment for our testbed is constructed on the basis of openstack technology. Then we utilize Spring Framework to implement the UH-BigDataSys system and define all services as restful API. Finally all kinds of clients access restful API to fetch all implemented services.

\begin{figure*}
\centering
\subfigure[]{\label{fig05a} \includegraphics[width=5.5in]{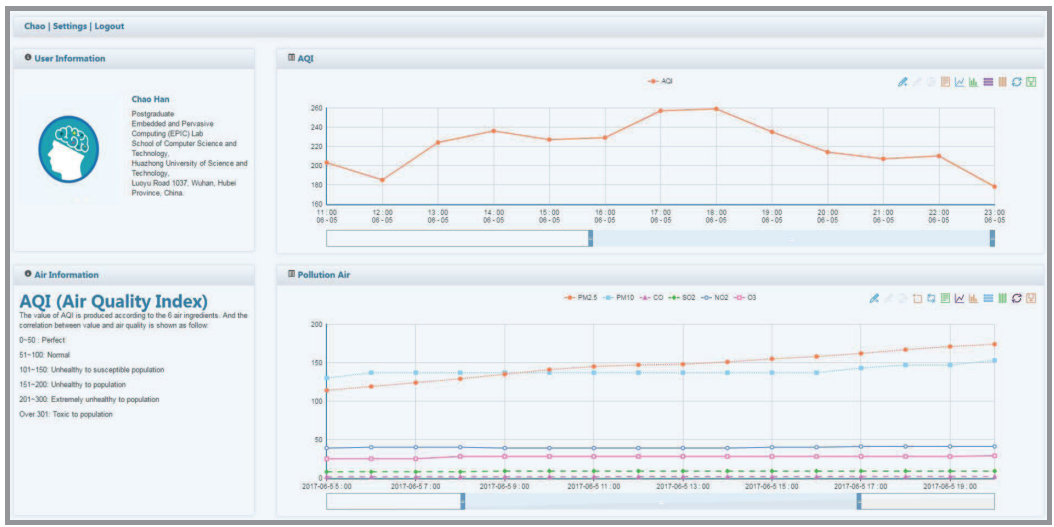}}
\subfigure[]{\label{fig05b} \includegraphics[width=5.5in]{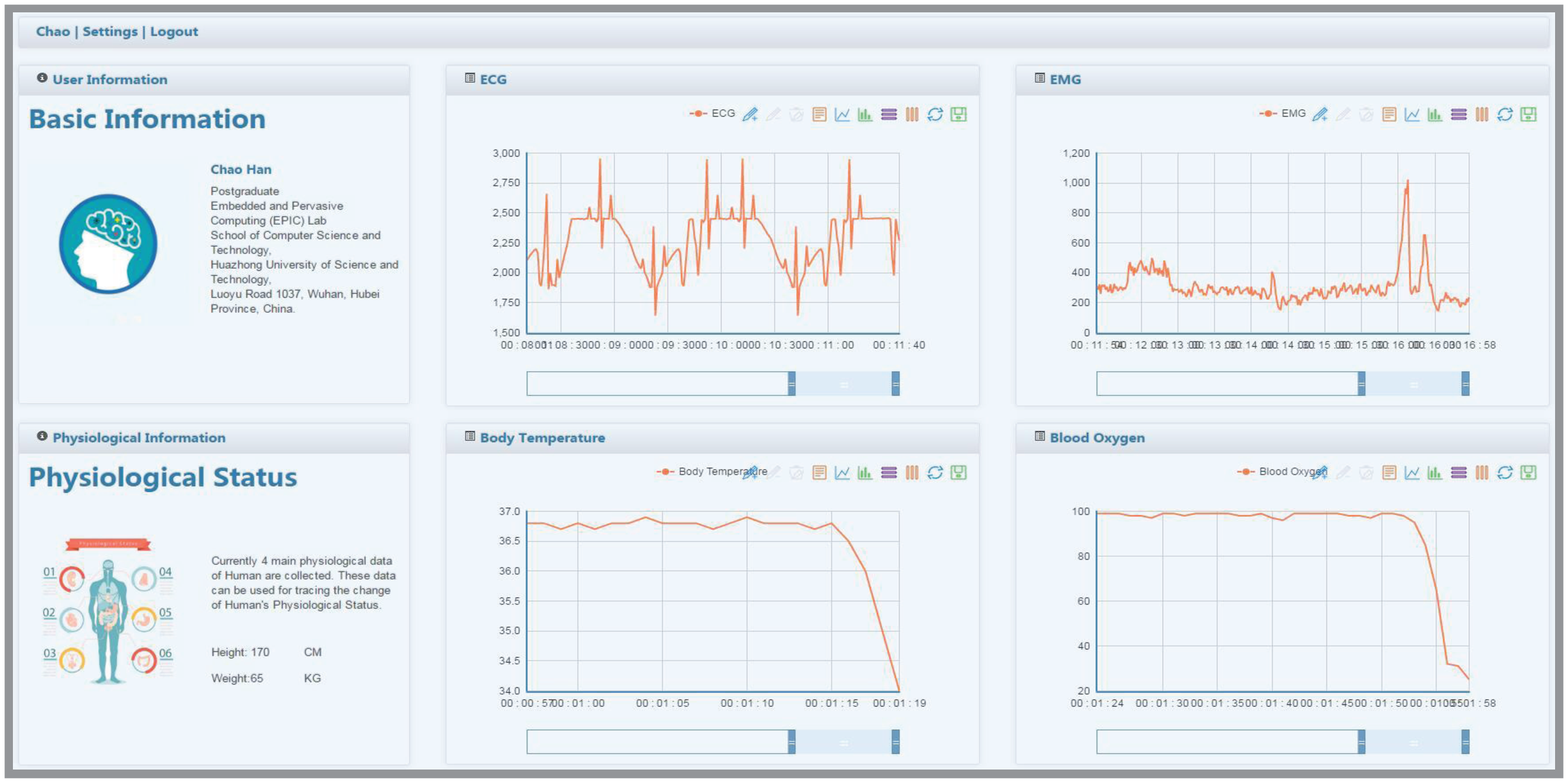}}
\caption{Web Interface of Testbed. (\textbf{a}) Air Quality Data, (\textbf{b}) Physiological Data.}
\label{fig005}
\end{figure*}

Fig.~\ref{fig004} and Fig.~\ref{fig005} exhibit the user interface of our UH-BigDataSys testbed.
Fig.~\ref{fig04b} and Fig.~\ref{fig05a} shows urban air quality data acquired by portable sensors carried by urban residents through mobile crowdsourcing. Fig.~\ref{fig04c} and Fig.~\ref{fig05b} shows physiological data of residents acquired by smart clothing. Fig.~\ref{fig04d} shows health advices given by UH-BigDataSys based on the acquired air quality around residents and the physiological status of residents.

As shown in Fig.~\ref{fig04a}, the sleep quality of a resident is closely related to his or her daily activities, surrounding air quality and his or her physiological and psychological state. UH-BigDataSys perceives physiological information of residents such as heart rate, blood oxygen, body temperature and exercise status via smart clothing, and obtains air quality data via crowdsourcing and IoT sensing, then utilizes big data analysis and machine learning to establish efficient prediction model, which guides behaviors of residents to improve their sleep quality. We found that psychological state, physiological state and sleep quality can affect each other.

\section{Conclusion}
In this article, we first discuss AQI data collection based on meteorological sites data, mobile crowdsourcing sensing and IoT sensing.
Then, the integration of M-AQI is proposed based on the edge clouds. M-AQI big data exhibits higher data quality and finer granularity. Next, physiological data of urban residents based on smart clothing is discussed and UH-BigDataSys is proposed. The system analyse M-AQI and physiological big data to provide guidance for urban residents in aspects such as respiratory disease, travel advice and sleep quality, in order to improve the life quality of urban residents. Finally, a tested of UH-BigDataSys is established towards the smart applications for enhanced healthcare based on AQI in urban environments.


\bibliographystyle{IEEEtran}

\end{document}